\documentclass[a4paper]{desyproc}
\usepackage{graphicx}
\begin{document}

\title{Solar flares as harbinger of new physics}

\author{K.~Zioutas$^1$,  M.~Tsagri$^2$\thanks{Part of the PhD thesis.}, Y.~Semertzidis$^3$, T.~Papaevangelou$^4$, E.~Georgiopoulou$^1$,
A.~Gardikiotis$^1$ and T.~Dafni$^5$\thanks{tdafni@unizar.es}\\ \\ 
 $^1$Physics Department, University of Patras, Patras, Greece\\
 $^2$European Organization for Nuclear Research (CERN), Gen\`eve, Switzerland\\
 $^3$Brookhaven National Laboratory, NY, USA\\
 $^4$IRFU, Centre d'\'Etudes Nucl\'eaires de Saclay (CEA-Saclay), Gif-sur-Yvette, France \\
 $^5$Grupo de F\'isica Nuclear y Astropart\'iculas, Universidad de Zaragoza, Zaragoza, Spain
} 

\maketitle

\begin{abstract}
The trigger mechanism of the energy release of solar flares is still unknown. This work provides additional evidence on the involvement of exotic particles like axions and/or other WISPs. The axion scenario suggests a dynamical behaviour behind white-light (WL) solar flares, which results to the measured spectral shape, and, the timing between visible-UV light and soft X-rays. Similarly, axions converted some 1000 km inside the photosphere with a rest mass of $\sim$17$\,$meV fit the unexpected soft X-ray emission from the quietest Sun in 2009, i.e., the manifestation of the solar corona problem. Chameleons behave similarly, but they couple to photons enhanced in vacuum, i.e., in the magnetized empty outer space, since their coherent conversion length stretches with better vacuum.
\end{abstract}

\section{Introduction}
The solar coronal heating mechanism remains an unsolved problem since 1939, being therefore one of the prominent challenges in solar physics and astrophysics. In addition, the so called White-Light (WL) solar flares were observed for the first time in 1859, but their trigger remains elusive, and ``flares might be unpredictable for more fundamental reasons" \cite{fundamental}. In this work we provide additional evidence that the working principle of CAST can be at work at the Sun, where outstreaming solar axions can convert to the otherwise unexpectedly measured X-rays \cite{njp}. If CAST's working principle has been fine tuned by nature at the Sun, then the most promising places to look out for similar effects, should be preferentially above Active Regions (ARs), i.e., sunspots. Note that these regions are also the birth places of flares, while the corona above non-flaring/quiet ARs is hotter than the rest of the quiet Corona. The temperature reaches $\sim$10$\,$MK and the unexpectedly emitted photons have a characteristic spectral shape (power law) that resembles down-comptonized hard X-rays appearing otherwise unexpectedly \cite{njp}. Indeed, the ``{\it Sun's intense X-ray emission is a remarkable and fascinating...mystery}"\cite{tsuneta}.
The idea of the down-comptonization underneath magnetized solar surface has been presented in \cite{njp}. Below, we update some of the conclusions in \cite{njp}, following recent X-ray measurements with SPHINX  in the $\sim$1--5$\,$keV.

\section{The new observational evidence}
\subsection{The X-ray emission from the quietest Sun in 2009}

\begin{figure}[hb]
\includegraphics[width=0.6\textwidth]{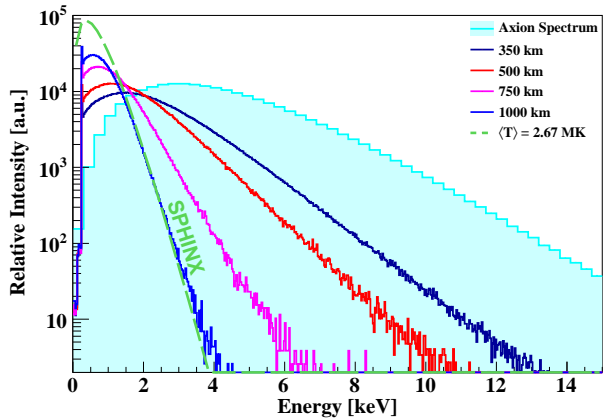}\hspace{0.5pc}%
\begin{minipage}[b]{13pc}\caption{The shadowed histogram gives the expected converted solar axion spectrum. Four degraded spectra due to multiple Compton scattering from 4 different depths into the photosphere are also shown. The spectrum measured by the SPHINX mission (green dashed line)  during the extremely quiet Sun in 2009 \cite{gburek1} agrees with that of the axion or axion-like scenario, assuming the conversion place is ~1000$\,$km underneath the photosphere \cite{njp}.}\label{figure1}
\end{minipage}
\end{figure}

The X-ray mission SPHINX has measured a minimum basal level of solar X-ray emission even during the most quiet solar minimum without flares or sunspots in 2009, below which solar activity never dropped. The measured quiet solar X-ray luminosity is $L_x\approx$ 6.7$\times$10$^{20}\,$erg/s$\,\approx$ 1.5$\times$10$^{-13}$$\,$$L_{\odot}$ \cite{gburek}. This fraction of the total solar luminosity is even 10 times below the estimated axion-photon conversion of about 10$^{-12}$ (see Footnotes 9 and 14 of \cite{njp}), making thus a quantitative X-ray intensity reconstruction reasonable, assuming that axions escape from the solar hot core with a rest mass of about 17$\,$meV and do convert to X-rays near the magnetized solar surface. In addition, the spectral shape follows a perfect power-law as it is shown in Figure \ref{figure1}, pointing at a depth of 1000$\,$km below the surface as the initiation place of the following down-comptonization. The spectral agreement between observation (SPHINX) and simulation extends over 3-4 orders of magnitude (see Figure \ref{figure1}).
The soft X-ray component below $\sim$2.5$\,$keV could be explained in general with light axions or other particles with similar couplings \cite{njp}. An alternative scenario is that with the solar chameleons \cite{cham,cham2}. In fact, X-ray photons from regenerated outstreaming chameleons could emerge in the outer solar magnetic fields, resembling that of the 2-3 MK quiet Sun X-ray emission (see Figure \ref{figure1}).  If the rest of the spectrum above 2.5$\,$keV is real \cite{gburek1}, such a second component might come, in principle, from regenerated axions in the upper photosphere / lower atmosphere, with suppressed down-comptonization. Furthermore, other exotica like radiatively decaying massive particles \cite{luigi} could also be at the origin.

%

\subsection{The White-Light flares: analog spectrum and time correlations}
\begin{figure}[hb]
\includegraphics[width=0.58\textwidth]{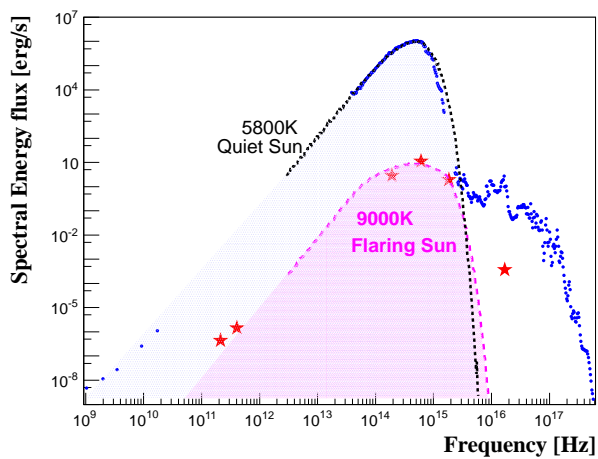}\hspace{0.5pc}%
\begin{minipage}[b]{13.5pc}\caption{The continuum flare spectrum
(red stars), with the dashed line representing a blackbody at $\sim$9000$\,$K
For comparison, it is also given the spectrum of the
quiet Sun (blue points). The overall spectrum of the quiet Sun has a near blackbody shape at visible and near-IR. Excesses over a
fitted blackbody appear at EUV and far-IR  and beyond. The striking EUV excess of the quiet Sun is the manifestation of the solar corona problem.
Courtesy H.S. Hudson \cite{hudson}.}\label{figure2}
\end{minipage}
\end{figure}

Figure \ref{figure2} shows the quiet Sun analog spectrum (in blue) \cite{hudson} along with a few points (stars in red) of the white-light (WL) emission during  a flare. One notes at the high-energy end the one flare-related point, which is far above that expected from the $\sim$9000$\,$K WL flare black-body emission. Thus, comparing with the quiet Sun spectral distribution with its striking photon excess at the high energy end ($>10^{15}\,$Hz), which reflects the solar corona problem, it seems that a flare develops its own and relatively more intense ``corona". This flare related ``corona" is even better seen in Figure 2 of ref. \cite{rhessi}, where the EUV intensity resembles also a flare's intense ``corona", being far above that of a black body WL flare distribution of $\sim 10^4$K. The appearance of a flare ``corona" fits our axion or axion-like picture, since it is widely accepted that a flare is magnetic in origin, though with several open questions remaining, like: where, when and how electrons are accelerated \cite{fleishman}? In the axion scenario an intervening magnetic field is just the catalyst, which transforms outstreaming axions or other particles like chameleons \cite{cham,cham2} to photons near and/or far from the Sun. As it was noticed in \cite{kretz,kretz2}, the magnetic field is a viable flare forecasting tool, while what powers/triggers a solar eruption is not known. But, ``\emph{it might be possible that an unknown mechanism produces the black body white-light flare spectrum near 9000$\,$K}" \cite{kretz,kretz2}(Figure \ref{figure2}). Moreover, the total energy radiated by flares exceeds by $\sim$100$\times$ the flare soft X-ray energy emission, with a major contribution in the visible and near-UV. Then, the required axion conversion efficiency is not unnaturally large.

\begin{figure}[hb]
\includegraphics[width=0.58\textwidth]{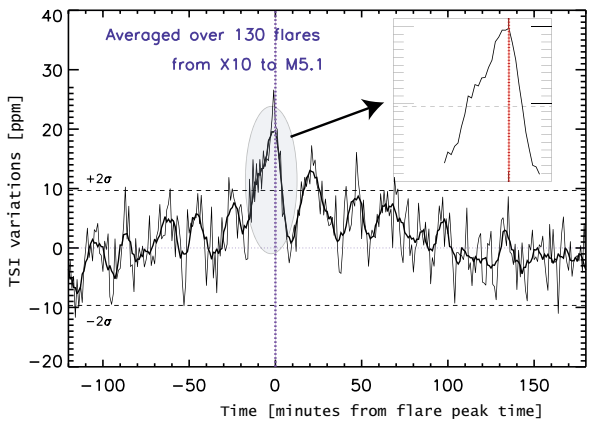}\hspace{0.5pc}%
\begin{minipage}[b]{14pc}\caption{The excess time distribution of the Total Solar Irradiance (TSI) emission of 130 flares relative to the soft X-ray peak (= zero time) \cite{kretz,kretz2}. X-rays come about $\,$5 minutes later than the extra light emitted during flares. Courtesy M. Kretzschmar.}
\label{figure3}
\end{minipage}
\end{figure}

As it is shown in Figure \ref{figure3}, the visible light (TSI) appears first and \emph{later} takes place the soft X-ray emission.
The axion scenario \cite{njp} for the WL flares is this: the energy deposition from converted axions streaming out of the inner Sun occurs at some deep photospheric layer. This gives rise to local ionization and thermalization, once a sufficiently strong magnetic field is combined with the appropriate density (which fits an axion rest mass of about 17$\,$meV \cite{njp}), maximizing thus the coherence length for the axion-to-photon conversion to happen efficiently. The so heated-up environment does not  thermalize completely to $\sim$6000$\,$K, and photons are escaping from the photosphere being a little hotter ($\sim$ 9000 K).
In addition, thanks to the horizontal magnetic field component, an outwards moving compressed front is possible, which pushes the resonant axion-conversion place ($\rho\approx m_{axion}$) upwards. This results in a decreasing column density above the actual conversion place. With time, the newly back-converted hard X-rays  1) do propagate in an ionized environment, and 2) suffer a more and more limited down-comptonization until they can leave the Sun as soft X-rays \cite{njp}.

In other words, within the axion scenario, the sequence of events, is this:
first starts the energy deposition by axion conversion in a relatively deep photospheric layer; this gives rise to the ambient ionization associated with a rather incomplete thermalization of about 10$^4\,$K with the escaping WL. The initial hard X-rays make their way outwards in a random walk, once the ambient plasma is complete, which allows multiple Compton scatterings to occur. Following this reasoning, the soft X-ray emission can come only later, and this is what has been observed recently, but only (!) when analyzing many flares \cite{kretz,kretz2}. An estimate of this time delay is interesting: the outward propagation speed of any energy deposition \cite{angelos}, e.g., by converted axions or other exotica, is about 1$\,$km/s;  this determines the compression time of the moving heated layer by some 500$\,$km upwards inside the photosphere, which takes some 10 minutes. Interestingly, this order of magnitude estimate fits the  observed delay of $\sim$5 minutes (see Figure \ref{figure3}). In summary:
\begin{itemize}
  \item [a.] the WL can come from not completely thermalized $\sim \,$4$\,$keV photons from converted axions (inverse Primakoff effect) at large photospheric depths, and
  \item [b.] this additional radiation pressure combined with the magnetic field in sub-photospheric layers \cite{schri} pushes the ionized axion conversion place upwards; the decreasing depth of the axion conversion layer (relative to that of the WL source origin), allows the X-rays to undergo a limited down-comptonization before escaping as soft X-rays with a characteristic power-law spectral shape, and this occurs $\,$5 minutes later, matching also observation.
\end{itemize}

Thus, a chain of processes observed during WL flares fit the axion scenario. However, this does not exclude synergism with conventional physics reactions and/or the involvement of other WISPs; axions \cite{njp} and chameleons \cite{cham,cham2} are rather two generic examples, which can be at work underneath and above the solar surface, respectively. Most probably, the conventionally mysterious multifaceted Sun cannot be understood only by a single global reaction mechanism.

\section{Discussion}

Outstreaming solar axions with a rest mass of $\sim$17$\,$meV can explain solar X-ray activity being enhanced above the magnetized photosphere \cite{njp}. Here we have elaborated the underlying processes, following the recent findings of WL flares. After the same reasoning, the quiet and the flaring Sun appear as two extreme cases. Within the suggested axion scenario, and assuming that the same chain of processes happens in both cases, the photosphere dynamics determines the spectral outcome. Because, the initial axion conversion layer (i.e., the origin of the WL) can move with the compressed hot front at less deep layers upwards, being thus shielded with a decreased column density above (i.e., the origin of the X-ray emission). This defines also the degree of the down-comptonization of the escaping radiation,  --as it was pointed out in Figure 10 of \cite{njp}-- the solar magnetic fields appear deeper in the quiet Sun than in the ARs, and therefore the quiet Sun is less X-ray active as a decreasing number of X-rays can reach the surface from deeper layers. This fits the fact that the solar corona above non-flaring ARs is hotter than that of the near quiet Sun ($\sim$10$\,$MK {\it vs.} $\sim$2$\,$MK) \cite{njp}. While any approach must explain first why the quiet corona is hot at all, before asking: why is it hotter there? Though, corona's dynamical behaviour might point at its workings.

As a second example in support of potential exotic particle involvement we mention solar chameleons with an energy of about 600$\,$eV, which can be created near the strongly magnetized tachocline \cite{cham2}. Since their conversion efficiency is maximum in vacuum, the magnetized solar outer space appears as their favourable place to get back-converted to soft X-rays by the inverse Primakoff effect. For example, the strong dipole magnetic field component between Sun and Earth (\textbf{BL} $\approx$10$^5\,$Tm) is of relevance only for particles like chameleons. Similar or even stronger fields appear in Corona Mass Ejections  (\textbf{BL} $\approx$ 10$^6\,$Tm), since the magnetic field strength is up to 200$\,$Gauss \cite{mittal}. Note, CAST has {\bf BL}=84$\,$Tm, while the signal strength goes with ({\bf BL})$^2$. Then, the Sun, its near and/or far transverse magnetic field components as seen from an orbiting observatory, being X-ray sensitive up to 10$\,$keV, resemble an Earth bound axion helioscope like CAST or Sumico. Therefore, such a ``natural" helioscope might have occasionally an enormous built-in axion or chameleon parameter fine tuning, which man-made equipment did not yet have the time to reach.

\section{Acknowledgments}
We are thankful to  Klaus Galsgaard, Manolis Georgoulis and Angelos Vourlidas for interesting discussions and suggestions. We also thank Hugh Hudson for allowing us to use Figure 2, and, Matthieu Kretzschmar  for kindly generating for us Figure 3.


\begin{footnotesize}

\end{footnotesize}


\end{document}